\documentclass[%
 twocolumn,
superscriptaddress,
 amsmath,amssymb,
prx,
longbibliography
]{revtex4-2}
\usepackage[utf8]{inputenc}
\usepackage{amsmath}
\usepackage{amssymb}
\usepackage{amsbsy}
\usepackage{amsthm}
\usepackage{mathtools}
\usepackage{xcolor}
\usepackage{thmtools}
\usepackage{thm-restate}
\usepackage{algorithm}
\usepackage{algorithmic}
\usepackage{braket}
\usepackage{acronym}
\usepackage{hyperref}
\hypersetup{colorlinks=true, citecolor=blue, urlcolor=blue, linkcolor=blue}

\newacro{AE}{Amplitude Estimation}
\newacro{QPE}{Quantum Phase Estimation}

\makeatletter
\newcommand{\dotDelta}{{\vphantom{\Delta}\mathpalette\d@tD@lta\relax}}
\newcommand{\d@tD@lta}[2]{%
	\ooalign{\hidewidth$\m@th#1\mkern-1mu\cdot$\hidewidth\cr$\m@th#1\Delta$\cr}%
}
\makeatother

\newcommand{\C}{\mathbb{C}}

\newcommand{\R}{\mathbb{R}}

\newcommand{\beq}{\begin{equation}}
	\newcommand{\eeq}{\end{equation}}
\newcommand{\beqn}{\begin{equation*}}
	\newcommand{\eeqn}{\end{equation*}}
\newcommand{\beqr}{\begin{eqnarray}}
	\newcommand{\eeqr}{\end{eqnarray}}
\newcommand{\beqrn}{\begin{eqnarray*}}
	\newcommand{\eeqrn}{\end{eqnarray*}}
\newcommand{\bmline}{\begin{multline}}
	\newcommand{\emline}{\end{multline}}
\newcommand{\bmlinen}{\begin{multline*}}
	\newcommand{\emlinen}{\end{multline*}}

\newtheorem{theorem}{Theorem}[section]

\newtheorem{definition}[theorem]{Definition}

\renewcommand{\ell}{L}

\begin{document}

\title{Quantum amplitude estimation from classical signal processing}

\author{Farrokh Labib}\email{farrokh@unitary.foundation}\affiliation{Unitary Foundation}
\author{B. David Clader}\affiliation{BQP Advisors, LLC, Ellicott City, MD}
\author{Nikitas Stamatopoulos}\affiliation{Goldman Sachs, New York, NY}
\author{William J. Zeng}\affiliation{Unitary Foundation}
\affiliation{Quantonation}

\begin{abstract}
    We demonstrate that the problem of amplitude estimation, a core subroutine used in many quantum algorithms, can be mapped directly to a problem in signal processing called direction of arrival (DOA) estimation. The DOA task is to determine the direction of arrival of an incoming wave with the fewest possible measurements. The connection between amplitude estimation and DOA allows us to make use of the vast amount of signal processing algorithms to post-process the measurements of the Grover iterator at predefined depths. Using an off-the-shelf DOA algorithm called ESPRIT together with a compressed-sensing based sampling approach, we create a phase-estimation free, parallel quantum amplitude estimation (QAE) algorithm with a worst-case sequential query complexity of $\sim 4.3/\varepsilon$ and a parallel query complexity of $\sim 0.26/\varepsilon$ at 95\% confidence. This performance is statistically equivalent and a $16\times$ improvement over Rall and Fuller~\href{https://doi.org/10.22331/q-2023-03-02-937}{[Quantum 7, 937 (2023)]}, for sequential and parallel query complexity respectively, which to our knowledge is the best published result for amplitude estimation. 
    The approach presented here provides a simple, robust, parallel method to performing QAE, with many possible avenues for improvement borrowing ideas from the wealth of literature in classical signal processing.
\end{abstract}

 \maketitle
	
\section{Introduction}
\ac{AE} \cite{brassard2002quantum} is a fundamental quantum algorithm with many applications. For example, it provides a quadratic speedup in Monte Carlo methods \cite{montanaro2015quantum}, giving speedups to problems in the financial sector \cite{rebentrost2018quantum, Stamatopoulos2020optionpricingusing, Chakrabarti2021thresholdquantum, Stamatopoulos2022towardsquantum}. AE is also a subroutine used to improve the complexity of algorithms that must estimate, for example, overlaps of states at the end of the quantum linear system algorithm \cite{PhysRevLett.103.150502, PhysRevLett.110.250504}.

\ac{AE} was first introduced as a combination of Grover search \cite{Grover1996} and \ac{QPE} \cite{nielsen2010quantum}. However, it was long conjectured that \ac{QPE} was not necessary since it typically provides exponential speedup, while \ac{AE} only provides a quadratic speedup over classical algorithms. This was proven true when Suzuki \textit{et al.,} showed that using Grover's algorithm combined with classical maximum likelihood estimation based post-processing achieved the optimal scaling without requiring \ac{QPE} \cite{suzuki2020amplitude}. This approach has been improved \cite{aaronson2020quantum, venkateswaran2020quantum, grinko2021iterative} with the result that achieves the best known query complexity, to our knowledge, being the one based on quantum signal processing \cite{Rall2023amplitudeestimation}.
  
Quantum algorithms for \ac{AE} without \ac{QPE} take measurements of the quantum state at different values of $n$, the number of applications of the Grover iterator, and use classical post-processing either at the end \cite{venkateswaran2020quantum,giurgica2022low, Rall2023amplitudeestimation} or iteratively \cite{aaronson2020quantum, grinko2021iterative} to determine at what $n$ to take samples next. The downside to the iterative approaches, as was pointed out in Refs. \cite{suzuki2020amplitude, venkateswaran2020quantum} is that one has to switch between quantum and classical repetitions in a serial manner, which may be undesirable in practice.

Here, we demonstrate a non-iterative \ac{AE} algorithm with the benefit that the number of iterations is known from the outset, so every sample can be done in parallel. In addition, the classical post-processing is a) robust to noise, allowing one to take very few samples to achieve low overall query complexity, and b) is classically efficient, not affecting the overall algorithm complexity up to log factors. Finally, in comparison to other similarly performant QAE algorithms, our approach does not require quantum signal processing (QSP) which simplifies the fault-tolerant circuits required to implement it. To achieve these results, we draw from ideas in (classical) signal processing, in particular from algorithms used in determining the Direction Of Arrival (DOA) of an incoming signal.

In DOA a set of sensors is placed at certain positions in space to detect an incoming signal from an unknown position \cite{krim1996, stoica2005spectral, theodoridis2013, Yang2018reivew, Gamba2019RadarSignalProcessing}. The measurement data is then used to determine the angle of arrival of the incoming signal relative to the position of the sensors. This problem has been well-studied as it has numerous applications in fields such as radar, sonar, and wireless communication. There are various algorithms for doing this, for example the MUltiple SIgnal Classification (MUSIC) algorithm \cite{barabell1998performance} or Estimation of Signal Parameters via Rotational Invariance Techniques (ESPRIT) \cite{roy1989esprit} are from a class of subspace methods, but there are many more variants and sampling techniques \cite{Yang2018reivew}.

The number of sensors and the spacing determines the accuracy or resolution with which one could estimate the DOA. Minimizing the number of sensors while simultaneously achieving high accuracy is desired for high-performance. It was discovered that using compressed-sensing methods allowed one to create what were called virtual arrays from sparse arrays, where the effective sensor spacing was greater than the number of actual sensors. 

One example are coprime arrays, where one combines two uniform samplers with sample spacings $MT$ and $NT$ where $M$ and $N$ are coprime integers and $T$ has dimension of space or time. This allows one to generate $\mathcal{O}(MN)$ sample locations using only $\mathcal{O}(M+N)$ physical samples \cite{pal2011coprime}. This can be further improved using multiple level nested arrays to achieve $\mathcal{O}(N^{2q})$ virtual sensors for some integer $q$ using just $\mathcal{O}(N)$ physical sensors \cite{pal2011multiple}.

It turns out that there is a nearly one-to-one correspondence between the DOA estimation problem and the amplitude estimation problem. This allows us to convert quantum measurements into a signal that can be post-processed using these DOA algorithms to estimate the amplitude. Our results demonstrate just one type of sensor spacing approach, the $2q$ array \cite{pal2011multiple} and one post-processing algorithm ESPRIT \cite{roy1989esprit}. We note that there are a wide number of variations of different sampling strategies and post-processing algorithms that could be further explored to optimize this approach to particular quantum algorithm and application needs \cite{Yang2018reivew}.

\section{Classical signal processing based AE}\label{section:csAE}
The setting is as follows: suppose we have access to a unitary $U$ such that 
\begin{align}
\label{eq:uop}
    U\ket{0^l} & = \cos\theta \ket{x,0} + \sin\theta \ket{x',1} \\ \nonumber
    & = \sqrt{1-a^2}\ket{x,0} + a \ket{x',1}
\end{align} 
for some $x,x'$ (states on $l-1$ qubits) and unknown $\theta\in[0,\pi/2]$. We wish to design an algorithm that finds $\theta$ (or the amplitude $a=\sin\theta$) up to an additive error $\varepsilon>0$. Classically, one would simply measure the state and conditioned on measuring a $\ket{0}$ or $\ket{1}$ in the final qubit and estimate $\theta$ using $\mathcal{O}(1/\varepsilon^2)$ samples, or applications of the unitary $U$. Using \ac{AE}, this can be improved to the optimal scaling that instead requires only $\mathcal{O}(1/\varepsilon)$ samples or applications of the unitary $U$, giving a quadratic speedup for quantum over classical.
	
To achieve the optimal scaling, let  $R_0$ be the reflection in $\ket{0^l}$, that is $R_0\ket{x} = -\ket{x}$ for $x\neq 0^l$ and $R_0\ket{0^l} = \ket{0^l}$. Let $S_0$ be the reflection in $\ket{0}$ in the last qubit, that is $S_0\ket{x,0}=\ket{x,0}$ and $S_0\ket{x,1}=-\ket{x,1}$ for all $x$. We can define the so-called Grover operator $G:=UR_0U^{-1}S_0$ which has the following property: after $n$ applications the state becomes
	\begin{align}\label{eq:grover_state}
		\ket{\phi_n} & := G^nU\ket{0^l} \\ \nonumber 
   & = \cos((2n+1)\theta)\ket{x,0}+\sin((2n+1)\theta)\ket{x',1}.
	\end{align}
When we measure the last qubit of the state $\ket{\phi_n}$ in the computational basis we obtain $\ket{0}$ or $\ket{1}$ with probabilities 
\begin{subequations}
\label{eq:zmeas}
	\begin{align}\label{eq:measurement_prob}
		p_0(n) & :=\cos^2((2n+1)\theta) \\
        p_1(n) & :=\sin^2((2n+1)\theta).
	\end{align}
\end{subequations} 
Using the double angle formula
\begin{equation}
	p_0(n) = \frac{1+\cos((2n+1)2\theta)}{2},
\end{equation}
we can obtain an estimate for $\cos((2n+1)2\theta)$ by taking repeated samples of the quantum state $\ket{\phi_n}$.

To apply signal processing techniques, we need to form the complex exponentials
\begin{equation}\label{eq:quantum_signal}
	y_n = e^{i [2\theta \pi (2n+1)]}.
\end{equation}
Using the double angle formula, we have access to estimates of the real part of $y_n$, but not the imaginary part. Hence, both $y_n$ and $y_n^*$ are consistent with the probabilities obtained from measuring the quantum state $\ket{\phi_n}$. We will show in Section~\ref{subsection:correct_signs} how to distinguish between the two cases using classical post-processing. For now, we assume we have access to estimates of both the real and imaginary parts of $y_n$.

The estimation of $y_n$ is not exact. The probabilities that we estimate from Eqs.~\eqref{eq:zmeas} have binomial sampling noise. The goal of \ac{AE} is to estimate $\theta$ using as few measurements as possible. In the next section, we describe a classical signal processing approach that uses estimates to Eq.~\eqref{eq:quantum_signal}, mimicking almost exactly the same type of estimator one obtains using physical sensors that are attempting to estimate the direction of arrival of an unknown signal using noisy measurements.
\subsection{Subspace Based Direction Of Arrival Estimation}
\begin{figure}[h]
\centering
\includegraphics[width = 0.97\columnwidth]{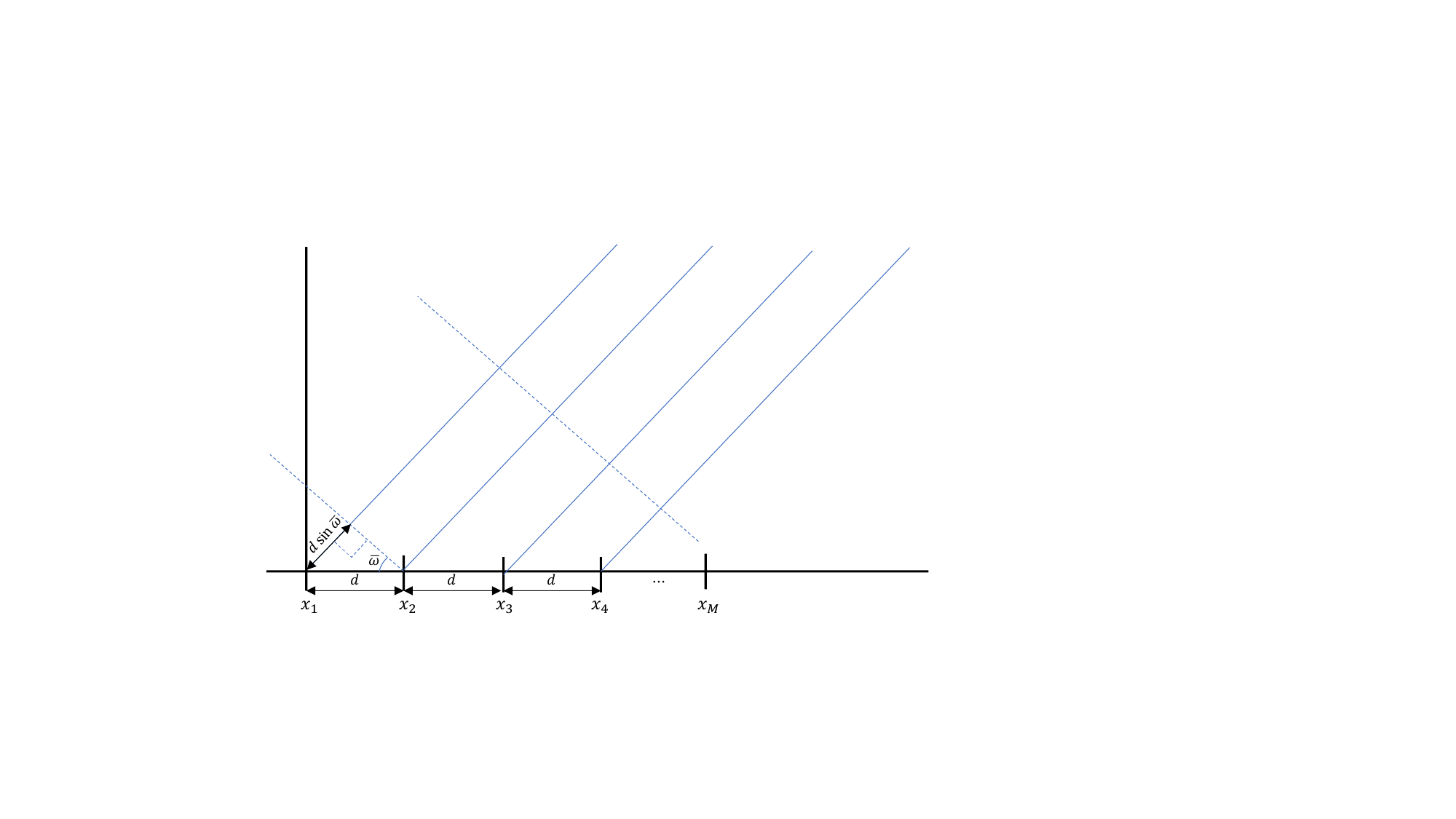}
\caption{\label{fig:sensor_geometry}Figure depicting the direction of arrival estimation geometry for a uniform linear array. Here a linear array of $M$ sensors is arrayed and the phase of the incident wavefront depends upon its angle of arrival.}
\end{figure}

Suppose a linear array of physical sensors placed at positions $x_1,\dots, x_M$ and suppose that there is one source providing an incoming signal with arrival angle $\bar\omega\in (-\pi/2,\pi/2)$ shown schematically in Fig. \ref{fig:sensor_geometry}. A sensor at position $x_n$ obtains a measurement of 
\begin{equation}
\label{eq:measurement_model}
y_n^\prime = e^{ i 2 \pi (n-1)d \sin \bar{\omega} /\lambda} + \epsilon_n = e^{i \omega x_n} + \epsilon_n = y_n + \epsilon_n,
\end{equation}
where we have identified a unit distance $x_n = (n-1)$ and angle $\omega = 2\pi d \sin{\bar{\omega}}/\lambda$ for an incoming signal with wavelength $\lambda$. The parameter $y_n$ here is distinct from that defined in Eq.~\eqref{eq:quantum_signal}, but we purposefully use the same notation as it turns out they are equivalent from the perspective of DOA estimation. The parameter $\epsilon_n$ denotes an error term due to imperfect measurements. The goal of DOA estimation, is to determine the incident angle of the incoming signal $\bar\omega$ using the fewest possible sensors. Looking at Eq.~\eqref{eq:quantum_signal}, we can view it as an incoming signal with angle of incidence $\omega = 4\theta$ at position $n$. This is the main connection between \ac{AE} and DOA estimation that we will use. The main difference is how we obtain noisy measurements of $y_n$.

Consider now the covariance matrix, $\boldsymbol{R}$, of the estimated signal vector $\boldsymbol{y}^\prime = (y_1^\prime, \ldots, y_M^\prime)^T$
\begin{equation}
    \label{eq:covariance}
    \boldsymbol {R} = E[\boldsymbol{y}^\prime \boldsymbol{y}^{\prime \dagger}] = \boldsymbol{y}\boldsymbol{y}^\dagger + \sigma^2 \boldsymbol{I},
\end{equation}
where $\sigma^2 = E[\epsilon_j, \epsilon_j]$ is the variance of the noise which is assumed to be Gaussian and uncorrelated, and $\boldsymbol{I}$ is the identity matrix. One of the key insight of subspace based DOA estimation algorithms, is that the covariance matrix can be decomposed into signal and noise subspaces, which can lead to accurate estimation of the DOA in the presence of noisy estimates. Consider the eigendecomposition of $\boldsymbol{R}$
\begin{equation}
    \label{eq:eigenR}
    \boldsymbol{R} \boldsymbol{u}_n = \left( \boldsymbol{y}\boldsymbol{y}^\dagger + \sigma ^2 \boldsymbol{I}\right) \boldsymbol{u}_n = \lambda_n \boldsymbol{u}_n.
\end{equation}
The eigenspace of the covariance matrix $\boldsymbol{R}$ can be partitioned into a signal subspace and a noise subspace. For \ac{AE} there is only one signal, so the signal subspace has dimension 1 with eigenvalue $\lambda_s = \sigma_s^2 + \sigma^2$ and eigenvector $\boldsymbol{u}_1$, while the noise subspace is size $M-1$ with eigenvalues $\lambda_n = \sigma^2$ and eigenvectors $\boldsymbol{u}_{n\ne 1}$.

It turns out that the eigenvectors associated with the noise subspace are orthogonal to the signal vector $\boldsymbol{y}$ \cite{Schmidt1986MUSIC}. That is, one can show that
\begin{equation}
    \label{eq:orthogonal}
    \boldsymbol{y}^\dagger\cdot \boldsymbol{u}_j = 0 \quad j \ne 1.
\end{equation}
This approach was one of the first subspace based approaches to DOA estimation, called the MUSIC algorithm. It estimates the angle of arrival of a signal, $\omega^\prime$, by finding angles where the signal subspace is orthogonal to the noise subspace \cite{Schmidt1986MUSIC}. The MUSIC algorithm performs well in practice, but it is overkill for what is required for \ac{AE}. It requires one to perform a full eigendecomposition of the $M\times M$ covariance matrix $\boldsymbol{R}$ which has complexity $\mathcal{O}(M^3)$ and also requires one to compute Eq.~\eqref{eq:orthogonal} for many different possible angles $\omega^\prime$ to achieve good performance. 

Standard DOA can handle multiple sources, but this is not necessary for \ac{AE}. However, we want to note that having the ability to handle multiple sources might be useful to optimize a core part in the quantum mean estimation algorithm of~\cite{kothari2023mean}. In their approach, the Grover operator that is used is not just 2-dimensional (in \ac{AE} the Grover operator is essentially a rotation on a plane), but can be high-dimensional. In signal processing terms, this corresponds to having multiple sources of direction of arrival.

\subsection{ESPRIT Approach to DOA Estimation}
Just like MUSIC, ESPRIT \cite{roy1989esprit} is an algorithm that uses the measurements $\boldsymbol{y}^\prime$ to estimate the unknown angle $\omega$. The difference is that instead of estimating the incoming spectrum using the noise eigenspace as shown in Eq.~\eqref{eq:orthogonal}, it just estimates the incoming DOA directly from the signal subspace. This allows a much more computationally efficient approach to DOA estimation. We outline this in Algorithm \ref{algo:ESPRIT} and refer the reader to Refs. \cite{roy1989esprit, liu2015spatialsmoothing, Gamba2019RadarSignalProcessing} for further details.



\begin{algorithm} [H]
\caption{ESPRIT}\label{algo:ESPRIT} 
\begin{algorithmic}[1]
\REQUIRE Measurements of a signal $y(x) = e^{i \omega x}$ on a uniformly spaced array of sensors.
\ENSURE An approximation of $\omega$ up to error $O(1/M)$ where $M$ is the size of the array of sensors.
\STATE Form the Toeplitz matrix \cite{liu2015spatialsmoothing} $\boldsymbol{R}$ with first row the vector of measurements $\boldsymbol{y}^T\in \C^M$ and first column the vector of measurements $\boldsymbol{y}^* \in \C^M$ and compute it's singular value decomposition $\boldsymbol{R} = \boldsymbol{U}\boldsymbol{\Sigma}\boldsymbol{V}$.
\STATE Form the matrix $\boldsymbol{S}$ from the first 2 columns of $\boldsymbol{U}$, the matrix $\boldsymbol{S}_1$ from the first $M-1$ rows of $\boldsymbol{S}$, and the matrix $\boldsymbol{S}_2$ from the last $M-1$ rows of $\boldsymbol{S}$.
\STATE Compute the $2\times 2$ matrix $\boldsymbol{P} = \boldsymbol{S}_1^{-1}\boldsymbol{S}_2$ where $\boldsymbol{S}_1^{-1}$ is the pseudo-inverse. 
\STATE Output the phase of the first eigenvalue of $\boldsymbol{P}$.
\end{algorithmic}
\end{algorithm}
For \ac{AE}, we don't need to do the full singular value decomposition of the matrix $\boldsymbol{R}$ because we know from the outset that there is just a single incoming signal. Instead, we just need the top two eigenvectors to form the matrices $\boldsymbol{S}_1$ and $\boldsymbol{S}_2$. Because the matrix is Toeplitz, we can use the Lanczos algorithm \cite{lanczos1950} together with Fourier-transform based algorithms for computing matrix-vector products of Toeplitz matrices to obtain these two eigenvectors in time $O(M\log(M))$ \cite{Golub1996}. Computing the pseudo-inverse of $\boldsymbol{S}_1$ is likewise fast, taking time $\mathcal{O}(M)$, since the matrix is of size $(M-1) \times 2$. Therefore the classical complexity of using ESPRIT for \ac{AE} is $\mathcal{O}(M\log(M))$, logarithmically worse than the lower-bound query complexity of \ac{AE}, but insignificant in practice.

\subsection{Physical and virtual arrays}
The previous descriptions, assumed a uniform array of sensors, denoted as a uniform linear array (ULA). However, this can be prohibitively costly as the required number of sensors can be large for good performance. A major advancement in this field was the development of the theory of sparse array processing \cite{Yang2018reivew} where one places sensors at only a few sub-sampled locations of the ULA which is called the physical array. The \textit{virtual array} concept is the idea that one can use the data from the physical array to get signal measurements at positions where there is no physical sensor, by combining the measurements of the signal from the physical array in a certain way. 

The ESPRIT algorithm, described previously, uses measurements of the incoming signal at physical positions which are evenly spaced. The virtual array concept allows us to obtain such measurements by using fewer physical sensors. As an example, suppose we place the sensors at physical locations $x = (x_1,x_2,\dots, x_n)$, which is a subset of sensor positions needed for the uniform linear array. Consider the vector of signals, $\boldsymbol{y}$, at those positions $y_j= e^{i \omega x_j}$. Compute the outer product of $y$ with itself, to obtain the following matrix
\begin{equation*}
(\boldsymbol{y}\boldsymbol{y}^\dagger)_{ij} = e^{ i \omega (x_i-x_j)}.
\end{equation*}
This is the value of the signal value at the physical location $x_i-x_j$. Suppose we have noisy measurements $y_j'$ of $y_j$ at locations $x_j$. We can use the value $(\boldsymbol{y}'\boldsymbol{y}'^{T})_{ij}$ to estimate the signal $\boldsymbol{y}$ at location $x_i-x_j$ even though we never physically measured the signal at that location. This technique allows us to obtain many more (virtual) measurements than the number of physical sensors. This approach is what allows us to achieve the optimal $\varepsilon \sim 1/N$ scaling for \ac{AE}.

Given a vector $x\in \R^n$ that represents a linear array of physical sensors where sensor $i$ is placed at position $x_i$, we can define for integer $q\geq 1$ its \textit{virtual array} by taking repeated outer products $q$ times.
\begin{definition}[Virtual array  \cite{chevalier1999virtual, Chevalier20062q}]
Let $x=(x_1,\dots, x_n)\in\R^n$ and $q\geq 1$ an integer. The $2q$-th order virtual array corresponding to $x$ is given by
\begin{equation}
S_q := \{ \sum_{i=1}^{q} x_{k_i} -\sum_{i=q+1}^{2q}x_{k_i} \colon k_i\in [n]\}.
\end{equation} 
\end{definition}

It is important to minimize the size of the physical array while maximizing the size of the virtual array. As we will see in the next section, the elements of the physical array $x_j$ corresponds to the number of Grover operators we apply.

We can create long virtual arrays using the following theorem on choosing the physical sensor locations \cite{pal2011multiple}.

\begin{theorem}\label{thm:nested_array}
Let $q, N_1,N_2,\dots, N_{2q}$ be positive integers and consider the following sets for $1\leq i\leq 2q-1$
\begin{equation}\label{eq:physical_array}
\{ n\prod_{k=0}^{i-1}N_k\colon n=1,2,\dots, N_i-1 \},
\end{equation}
and the $2q$-th set is given by $\{ n\prod_{k=0}^{2q-1}N_k\colon n=1,2,\dots, N_{2q} \}$ where $N_0:=1$.
Let the physical array be the union of all these sets.
Then the $2q$-th order virtual array corresponding to this physical array contains a ULA of size
\begin{equation}
2\prod_{k=0}^{2q}N_k -1.
\end{equation}
\end{theorem}

\begin{figure}[h]
\centering
\includegraphics[width = 0.99\columnwidth]{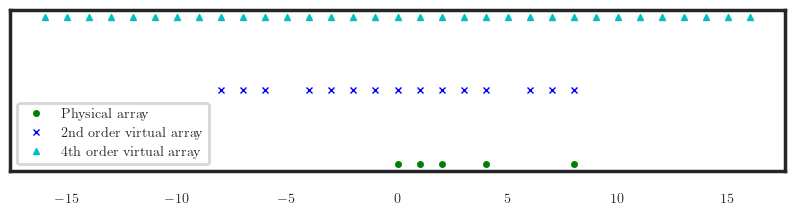}
\caption{\label{fig:virtual_array}Physical and virtual locations for $q=2$ and $N_1=N_2=N_3=N_4=2$. Note that the $2q$-th order (in this case 4th order) virtual array contains a large ULA.}
\end{figure}

As an example, we choose $N_1=N_2 =\dots = N_{2q}=2$ so that the union of all the sets in equation (\ref{eq:physical_array}) is equal to $(2^{j})_{j\in [2q]}$ and the corresponding virtual array has size $2^{2q+1}$. 
Fig.~\ref{fig:virtual_array} shows an example where $q=2$. The physical array is given by $\{1, 2, 4, 8\}$ which is shown by the green stars (also including the location 0). The second-order virtual array is the next level given by the dark blue dots, this is obtained by taking all the possible differences of pairs in the physical array. The 4-th order virtual array is given by the cyan triangles and it contains a ULA of size 31. Note that we only ever take measurements at the green dots, whose size (the number of locations) is significantly smaller than the virtual array (cyan dots) which is quantified by the above Theorem.

Once the virtual array is created using this method, we create what is known as a spatially smoothed version of the covariance matrix $\boldsymbol{R}$ from the virtual array using the method of Ref. \cite{liu2015spatialsmoothing}. This results in a Toeplitz covariance matrix that allows for efficient classical post-processing as remarked in the previous section.

Taking samples at depths of powers of two is reminiscent of the Maximum Likelihood Amplitude Estimation (MLAE) approach to AE~\cite{suzuki2020amplitude}. The main difference here is the classical post-processing, which provides far better results than MLAE. This was well-known for some time in the signal processing literature \cite{Gamba2019RadarSignalProcessing}. We used powers of two to determine scaling factors of our approach in Sec.~\ref{subsection:fits}, but note that there is a great deal of flexibility here depending on the use-case. 

\section{The AE algorithm}
\begin{algorithm} [H]
\caption{Amplitude Estimation from Classical Signal Processing (csAE)} \label{algo:AE} 
\begin{algorithmic}[1]
\REQUIRE Oracle $U$ such that $U\ket{0^l} = \cos\theta \ket{x,0} + \sin\theta \ket{x',1}$ and error rate $\varepsilon>0$.
\ENSURE Approximation of $a = \cos(\theta)$ to additive error $\varepsilon$.
\STATE Let $D$ be a set of depths such that the $2q$-th order virtual array has size $M\approx 1/\varepsilon$.
\STATE Measure the state $\ket{\phi_n}$ for $n\in D$ in the $Z$ basis and save the statistics $\hat{p_0}(n)$ ($\hat{p_0}(n)$ being the empirical probability of measuring $\ket{0}$).
\STATE Discretize the interval $[0,\pi/2]$ using $O(|D|)$ parts, learn the correct sign distribution and pick $\theta_{initial}$ in this discretization that is the most likely and use the signs of $\sin((2n+1)2\theta_{initial})$ as the initial guess, say $s^0=(s_1, \dots, s_{|D|})$.
\FOR{ $k$ in range($|D|$)}
\STATE Compute $S = \text{SlidingWindowSigns}(s^0, k)$.

\FOR{$s\in S$}
\STATE Form the ``signal'' vector $y_n = 2\hat{p_0}(n) - 1 + is_i\sqrt{1-[2\hat{p_0}(n) - 1]^2}$.
\STATE Form the signal vector $\boldsymbol{y}\in \C^M$ by computing the signal values at the virtual locations corresponding to positive virtual positions.
\STATE Use the ESPRIT (or any classical DOA) Algorithm~\ref{algo:ESPRIT} with $\boldsymbol{y}$ as input to extract the angle $\hat{\omega}$.
\STATE Compute its negative log likelihood and save the signs and estimated angle $\hat{\omega}$.
\ENDFOR

\ENDFOR
\STATE Return the angle $\hat{\omega}$ that minimized the objective function in the previous loop.
\end{algorithmic}
\end{algorithm}

We now show how these DOA algorithms outlined in the previous section can be used for \ac{AE}. Upon examination of Eq.~\eqref{eq:quantum_signal} and Eq.~\eqref{eq:measurement_model}, we see that the measurement model in the two approaches are identical and the DOA estimation methods are blind to the extra phase factor $e^{2i\theta}$ in $y_n$, meaning it will extract just $4\theta$ from measurements of $y_n$. 

In classical signal processing terms, we obtain an estimate of the of the signal $y_n =  e^{ i n 4\theta+2 i \theta}$ by placing a physical sensor at location $n$. Quantumly, we can obtain a estimate of the real-part of the signal $y_n$ by taking repeated measurements of $\ket{\phi_n}$ in the $Z$ basis. As far as the DOA algorithms are concerned it does not matter how the measurements are obtained. However, as mentioned in Section~\ref{section:csAE}, we can't quite distinguish between $y_n$ and $y_n^*$ from these measurement, since we can't deduce the correct quadrant of $(2n+1)2\theta$: using the value of $\cos((2n+1)2\theta)$ we can only obtain the value of $|\sin((2n+1)2\theta)|$. It's essential to obtain the correct sign of $\sin((2n+1)2\theta)$ so we can form the signal $y_n$.

\subsection{Obtaining the correct signs}\label{subsection:correct_signs}
To obtain the correct sign of $\sin((2n+1)2\theta)$ for each $n$ (in your sequence of depths $D$), we need an extra subroutine that checks whether an assignment of signs is likely or not.
For this, we need an objective function that assigns a score to each assignment of signs. Then we try many different assignments of signs (according to some heuristic) and we minimize the given objective function and use the signs that minimize the objective function to form the signal $y_n$. The algorithm that we propose for this purpose works in practice as our numerics will show, but can likely be further improved.

Given a sign vector and the estimated probabilities $\hat{p}_0(n)$ from Eq. \eqref{eq:zmeas} that we obtain by taking $N_\textrm{shots}(n)$ shots at each depth $n$, we can estimate $\theta_{est}$ using the ESPIRIT algorithm. From this estimate of $\theta_{est}$, we can compute an updated estimate $\hat{p}_0(n)$. For each of the different sign combinations, we minimize the negative log likelihood function
\begin{equation}
    \label{eq:nlogl}
    \begin{split}
    -\log(\mathcal{L}) & = -\sum_{n\in D}\bigg\{N_0(n)\log(\hat{p}_0(n)) \\
    & + [N_0(n) - N_\textrm{shots}(n)]\log(1-\hat{p}_0(n))\bigg\},
    \end{split}
\end{equation}
where $D$ is the set of depths at which we sample, $N_0(n)$ are the number of measurements that resulted in a 0, and $N_\textrm{shots}(n)$ are the total number of shots taken at depth $n$.

We have to be careful not to brute force all possible signs as the classical complexity will become exponential in the depth. Instead, we ``locally'' perturb the signs using a sliding window of fixed length, say two, and try all possible signs in that window while keeping the rest fixed. Given a vector of signs $s=(s_1,\dots, s_N)$ and integer $k$, the subroutine $\text{SlidingWindowSigns}(s,k)$ computes all the four sign vectors by varying the signs in the $k$-th window of size two in $s$. This increases the classical complexity of post-processing by a logarithmic factor of $O(\log(M))$. The numerics reported in Section~\ref{sec:numerics} uses sliding windows of size five instead of two.

We still have to make a choice of starting vector of signs. The way we approach this is to first ``learn'' the signs for each angle $\theta \in \Theta\subset [0,\pi/2]$ where $\Theta$ is a discrete subset of angles. The starting vector of signs is then the one that minimizes the objective function in this set. The size of $\Theta$ can be taken to be $O(|D|)$ so that the classical complexity increases by another logarithmic factor of $O(\log(M))$.
\subsection{Circuit Depth and Samples}
The next step is to determine at what depths we take measurements. The DOA algorithms expects measurements on a uniform linear array of length $M$, that is measurements of $y_n$ for $n\in [M]$. One may obtain this by measuring $\ket{\phi_n}$ at all depths $n\in [M]$, with a query complexity of $\sum_{n=1}^{M}n=O(M^2)$. This does not provide us with the expected quantum scaling $O(1/M)$ of the error, but rather the scaling is $O(1/\sqrt{M})$ in this case. We avoid this problem by taking samples from a far smaller set of depths and use the virtual array concept to obtain (virtual) measurements at all depths $n\in [M]$.

More precisely, we apply Theorem (\ref{thm:nested_array}) and use the physical array given by the sequence $D=(2^{j})_{j\in [2q]}$. The query complexity of taking measurements of $\ket{\phi_n}$ for all $n\in D$ is $O(2^{2q+1})$, while the size of the ULA inside the virtual array is $2^{2q+1}$ by Theorem (\ref{thm:nested_array}). Since the DOA algorithms can estimate the incident angle with accuracy $O(1/M)$ if $M$ is the length of the ULA, this implies immediately the quantum scaling that we were looking for. A summary of the full \ac{AE} algorithm is given in Algorithm \ref{algo:AE}.

\section{Numerics}\label{sec:numerics}
\begin{figure}[h!]
\centering
\includegraphics[width = 0.97\columnwidth]{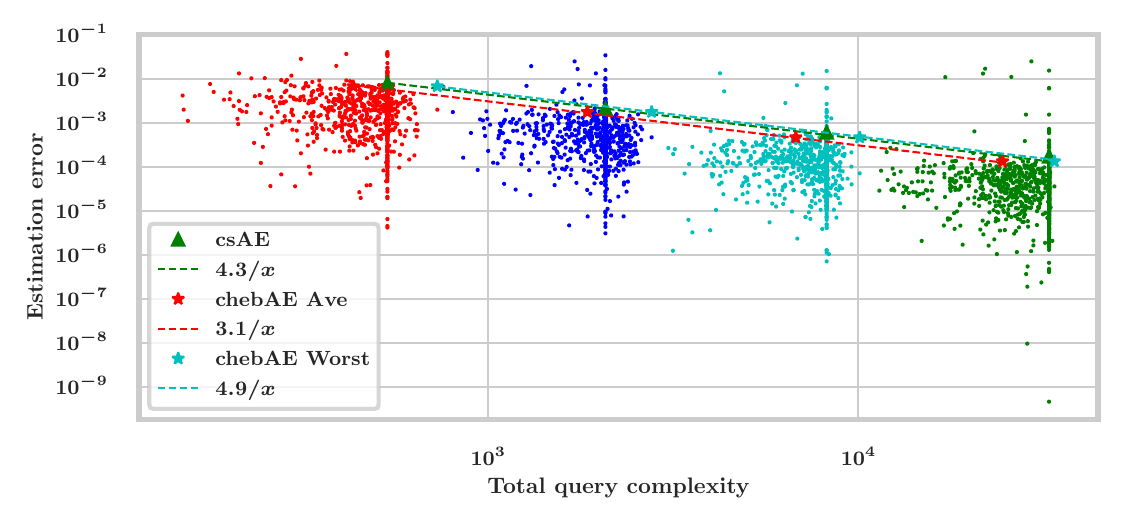}
\includegraphics[width = 0.97\columnwidth]{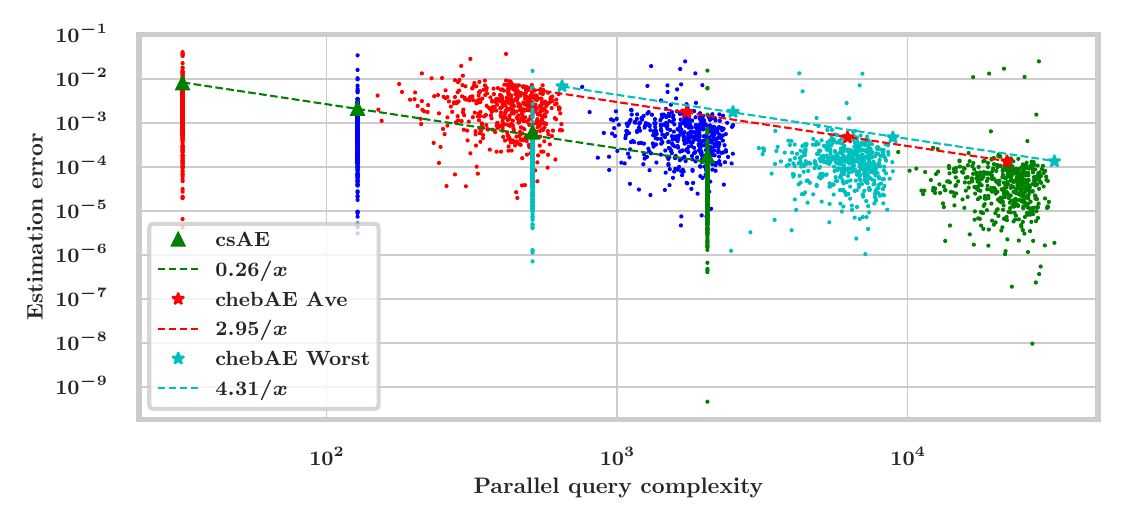}
\caption{\label{fig:query_complexity}Estimation error as a function of query complexity (top) and parallel complexity (bottom) for amplitude $a=0.5$. For csAE we simulated 500 runs with a sliding window value of length five giving a range of error rates (given by the vertical dots). The green triangles denote the 95-th percentile of the errors and the green fit line is used to compute the scaling factor. For chebAE we also simulated 500 runs, for a target $\varepsilon$ given by the results from csAE. The cloud of points from chebAE arise because of the iterative nature. The red and cyan stars and fit lines denote the average and worst case complexity respectively for chebAE for the 95-th percentile.}
\end{figure}

\begin{table*}[t]
\caption{\label{tab:my_table} Fitting constants for various confidence levels, along with the associated standard error obtained from the weighted-least-squares fit. The reported numbers are the average constant factors (resulting from the fit) on amplitudes sampled uniformly random in the interval $(0.1, 0.9)$. Overall, csAE is comparable to the average case complexity of chebAE in terms of total query complexity, but tends to outperform in terms of worst-case complexity. In addition, csAE is highly parallelizable resulting in vast out-performance for parallel query complexity.}
\begin{ruledtabular}
\begin{tabular}{ccccccc}
               & $99\%$ total  & $99\%$ parallel & $95\%$ total  & $95\%$ parallel   & $68\%$ total    & $68\%$ parallel \\
\hline
csAE           & $8.9 \pm 0.9$ & $0.27 \pm 0.03$ & $4.3 \pm 0.1$ & $0.26 \pm 0.01$ & $1.67 \pm 0.05$ & $0.102 \pm 0.003$ \\
chebAE (worst) & $6.1 \pm 0.2$ & $5.87 \pm 0.05$   & $4.88 \pm 0.05$ & $4.3 \pm 0.1$     & $2.05 \pm 0.02$ & $1.79 \pm 0.06$ \\
chebAE (ave)   & $ 4.1\pm 0.2$ & $3.9 \pm 0.2$   & $3.07 \pm 0.11$ & $2.95 \pm 0.06$     & $1.30 \pm 0.01$ & $1.19 \pm 0.02$ 
%
\end{tabular}
\end{ruledtabular}
\end{table*}


In this section we show the performance of our algorithm csAE and compare it against the state of the art Amplitude Estimation algorithm which we refer to as chebAE \cite{Rall2023amplitudeestimation}. The code for reproducing the numerics and plots in this paper can be found on~\href{https://github.com/unitaryfund/csAE}{https://github.com/unitaryfund/csAE}.

We compare the number of oracles queries required by the algorithm to estimate the amplitude $a$ defined in Eq. \eqref{eq:uop}. The metric we compare is the estimation error 
\begin{equation}
    \label{eq:estimation_error}
    \varepsilon = |a-\hat{a}|
\end{equation} 
where $\hat{a}$ is the estimated amplitude error rates of the algorithm from different runs. More formally, we also define a confidence level of our estimate. That is we estimate
\begin{equation}
    \textnormal{Pr}[|a-\hat{a}|<\varepsilon] \ge \delta
\end{equation}
where $\delta$ is the confidence level. 

The query complexity is the number of times we use the operator $U$ (see ~\eqref{eq:uop}) multiplied by two. Since the Grover operator $G$ consists of two calls to $U$, the query complexity will be the number of times we use $G$ plus the number of oracle calls at depth 0 divided by two. In chebAE~\cite{Rall2023amplitudeestimation} the number of oracle queries is defined as the number of times we apply the Grover operator $G$ ignoring the calls to $U$ at depth 0. This won't affect constant factors in any meaningful way (we are penalizing our algorithm slightly). We compare the two approaches ability to achieve a target $\varepsilon$ and $\delta$ on two different metrics: 1) query complexity and 2)  parallel query complexity. By parallel query complexity we mean the maximum depth oracle query that can be computed at compile time.

To get a better intuition of the performance of the algorithm, in Fig.~\ref{fig:query_complexity}, we provide results when estimating the amplitudes chosen uniformly at random over the range $a\in (0.1, 0.9)$ for 500 trials. We use a physical array with parameters $N_1 = N_2 = \cdots N_{2q} = 2$ as defined in Eq. \eqref{eq:physical_array} for $q \in \{3, 4, \ldots, 6\}$. For csAE, the input is the physical sampling array, denoting the fixed query depth, as well as the number of samples to take at each depth. This results in a deterministic schedule and yields the vertical colored dots representing the estimation errors of the algorithm from different runs. The green triangles represent the $\delta=0.95$ confidence level of those runs for different depth circuits. The green line is a fit to these 95\% confidence levels.

The number of samples to take at each depth is user-defined. We found that using a fixed number of shots was sub-optimal. The algorithm yielded better constant factor complexity when taking more shots at shorter depths and fewer shots at deeper circuit depths. As a heuristic, we used the following schedule
\begin{equation}
    \label{eq:nshots}
    N_{\mathrm{shots}}(n) = \lceil K(\log_2(n_{\mathrm{max}}/n) + 1)\rceil,
\end{equation}
where $n_{\mathrm{max}}$ is the maximum query depth, $n$ is the current query depth, and $K$ is a constant that we chose to be $K=4$ for $\delta \le 0.95$ and $K=8$ for $\delta = 0.99$. It is possible that there exist better schedules $N_{\mathrm{shots}}(n)$. We note that it is also possible to have a non-deterministic schedule that utilizes information from the the top-two eigenvalues of the matrix $\boldsymbol{P}$ of Algorithm \ref{algo:ESPRIT}. When these eigenvalues are close to the remaining eigenvalues the signal and noise subspaces are not well separated. We believe that using this information could further improve the constant factor scaling, but at the cost of reducing parallelizability. We leave this to future work.

We directly compare our results to the performance of chebAE in the same figure. In the case of chebAE, the estimation error and confidence-level are inputs to the algorithm while the query complexity is a random variable determined at runtime. This results in the cloud of points observed on the plot, which is the result of each independent run of chebAE. Because the query complexity is a random variable, we report both the $\delta=0.95$ confidence level for the average query complexity and worst case query complexity as the red and cyan points respectively. The red and cyan lines are fit lines to these 95\% confidence levels. In addition to the determinstic schedule benefit of csAE, we also note that the worst-case error rate is typically more constrained compared to chebAE, which can fail with much larger error rates for the parameters chosen.

\subsection{Fits}\label{subsection:fits}
To estimate the scaling factor, we perform a weighted least-squares fit to the function $N = C/\varepsilon_\delta + b$ ($N$ number of queries and $\varepsilon_\delta$ error rate at a given confidence $\delta$), where the weighting factor is given by $\varepsilon_\delta$. We get an estimate of the optimal parameter $C_{\mathrm{opt}}$ by minimizing the weighted residuals in the 2-norm: write $r_i := N_i-C/\varepsilon_{\delta, i}-b$ for the residual of observation $i$ and $w_i = \varepsilon_{\delta, i}$ the corresponding weight, then 
\begin{equation}
    \label{eq:cfit}
    C_{\mathrm{opt}}:= \min_C \sum_i w_ir_i^2.
\end{equation}
The performance of the algorithm depends on the value of the amplitude $a$ to be estimated. Since $a$ is an unknown parameter, we compare the performance of csAE for amplitudes sampled uniformly in the interval $(0.1, 0.9)$. For each amplitude, we perform the same analysis shown in Fig.~\ref{fig:query_complexity} and compute the estimation error $\varepsilon_\delta$ at a given confidence-level $\delta$. We compute the weighted-least squares fit using both average and worse-case query complexities for each amplitude to compute the scaling factor $C_{\textrm{opt}}$ and report the average number in Tab.~\ref{tab:my_table}. Here we provide a summary of the constant factor scaling comparing csAE to chebAE for different confidence levels and for both average and worst-case query complexity for chebAE. 

Overall, we find that our algorithm under-performs chebAE for the average case, but matches the performance in the worst case to within statistical uncertainties. The greatest benefit from using our algorithm is its parallelizability, which greatly outperforms chebAE since that algorithm is inherently sequential.


In addition to matching the performance of chebAE in terms of total oracle queries, the implementation of csAE is also simpler. As with the IQAE implementation \cite{grinko2021iterative}, the csAE approach makes use of the standard Grover iterator and does not require Quantum Signal Processing (QSP). This implies that the circuit-level implementation is simpler than for chebAE, not requiring controlled phase rotations and circuit synthesis, which adds a logarithmic factor to the required fault-tolerant circuit depth \cite{ross2016}. Therefore, our approach yields the best of both worlds: simple circuit construction, together with state-of-the-art query complexity. As we discuss next, our approach also offers a highly parallel implementation, something not available to either chebAE or IQAE.
\subsection{Target Error Performance}\label{subsection:target_error}
Since the query depths and number of samples are pre-determined for csAE (which enables its parallelizability), it requires some trial and error if one wishes instead to achieve a desired target estimation error $\epsilon_\delta$ for a given confidence level $\delta$. The fits provide a useful parameter with which to quickly estimate the number of queries, but here we also provide concrete examples of the number of queries required to achieve a target error of $\epsilon_\delta=10^{-3}$ with confidence levels of $99\%$, $95\%$ and $68\%$, which are often used in practice when performing detailed resource estimation.

Results are presented in Tab. \ref{tab:target_error}. As before, we estimate the performance by simulating the performance of the csAE algorithm for amplitudes sampled uniformly at random over the range $a=(0.1, 0.9)$ using 500 Monte Carlo trials. These results provide a quick reference with which one can perform detailed resource estimation for their desired application.

\begin{table}[t]
\caption{\label{tab:target_error} Sequential queries, maximum depth query, and simulated performance to achieve a target error rate of $\epsilon_\delta = 10^{-3}$ for various confidence levels $\delta$. Details of the array settings and sampling strategy are given in Tabs. \ref{tab:array_params99}-\ref{tab:array_params68} in App. \ref{app:target_error}.}
\begin{ruledtabular}
\begin{tabular}{ccccccc}
& Sequential Queries & Maximum Depth & Error\\
\hline
$\delta=0.99$ & 8,777 & 256 & $9.3\times 10^{-4}$ \\
$\delta=0.95$ & 4,488 & 256 & $8.0\times 10^{-4}$ \\
$\delta=0.68$ & 1,560 & 128 & $9.8\times 10^{-4}$ \\
\end{tabular}
\end{ruledtabular}
\end{table}

\subsection{Parallelizability}
\begin{figure}[h]
\centering
\includegraphics[width = 0.97\columnwidth]{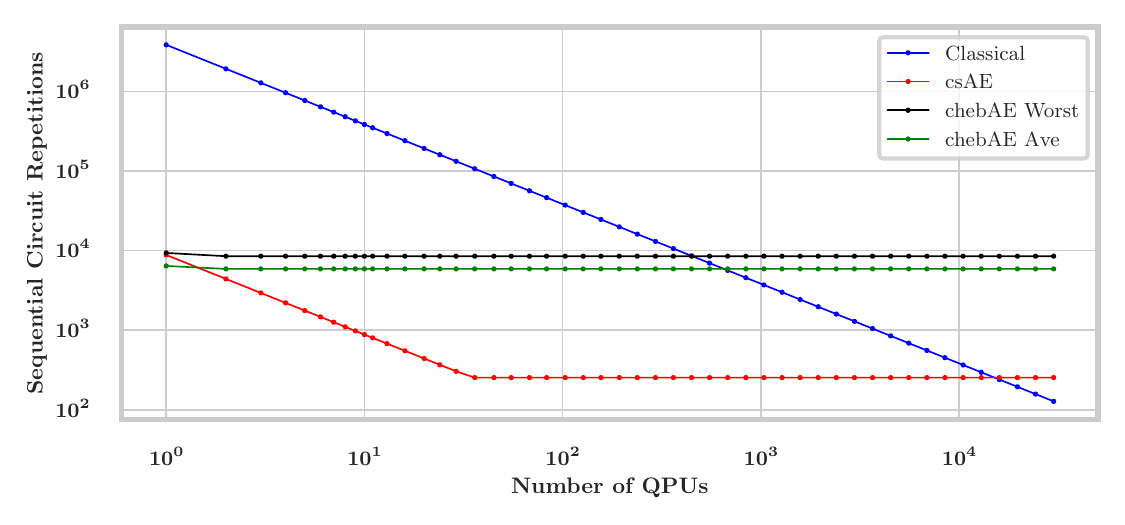}
\caption{\label{fig:constant_factor_plot}Required number of sequential applications of the circuit $U$ as we increase the number of quantum computers. This is for the amplitude $0.5$ to achieve the target error rate $\epsilon = 10^{-3}$ at 95\% confidence. The blue curve is for classical sampling where we simply apply the unitary $U$ once and sample. The black and green curves are for the worst and average case complexity of chebAE respectively. Here we assume at 2 QPUs chebAE can achieve maximum parallelizability, which is an optimistic assumption. Finally, the red curve is for csAE. Here we see the algorithm is fully parallel until it saturates. This occurs when the largest depth circuit is run on a single QPU for each shot. The remaining shorter depth circuits can be run somewhat sequentially on additional QPUs due to the exponential schedule.}
\end{figure}

The csAE algorithm is highly parallelizable, because the quantum circuits at which we take samples from is known beforehand. The results shown before provide the minimum depth circuit one can run assuming full parallelizability. Here we look to answer the question of how to parallelize the algorithm over a fixed number of QPUs. The best approach is to divide the total number of queries as evenly as possible among the available QPUs. The parallel query complexity is then simply the largest query complexity among the QPUs. We can find this even division using a greedy approach: sort the circuits from which we take samples in decreasing Grover power and distribute the circuits evenly among the QPUs starting with the highest Grover power.

In Fig.~\ref{fig:constant_factor_plot} we plot the maximum number of sequential applications of the circuit $U$ (which is twice the number of applications of the Grover oracles) given a number of QPUs assuming classical sampling, chebAE, and csAE. One can see that the parallelizability saturates for csAE, because the largest Grover power has to be run on some QPU, hence providing a lower bound on the parallel query complexity. Our results are presented assuming a target $\epsilon_{0.95}=10^{-3}$ consistent with row two of Tab. \ref{tab:target_error}.

In quantum computer architectures, where different Grover powers can be executed on the same QPU, the parallelizability of csAE also allows for the application of the technique known as Hamming weight phasing (HWP) which efficiently applies repeated arbitrary single-qubit rotations by the same angle \cite{Gidney_2018, Kivlichan_2020}. When $m$ repeated rotations by the same angle $R_z(\theta)$ are parallelizable, this technique reduces the number of rotations which need to be synthesized to $\lfloor \log_2 m +1 \rfloor$, at the cost of $m-1$ Toffoli gates and $m-1$ ancilla qubits. Because the Grover operator consists of repeated applications of a unitary $U$ (and $U^{\dagger}$), any single-qubit rotations in $U$ can therefore be applied using HWP across different powers of the Grover operator which have been parallelized on the QPU. The practical benefit from this approach depends on the specific form of $U$ and the total number of Grover powers that must be applied, but we highlight this ability to reduce the total number of single-qubit rotations across different Grover powers as this is not possible in iterative variants of AE.
\section{Discussion and Future Work}
We have shown that there is a direct correspondence between quantum amplitude estimation and the estimation of the direction of arrival of an incoming signal, a well-studied problem in classical signal-processing. This correspondence allowed us to provide a new \ac{AE} algorithm that is fully parallel and opens up the possibility of using the vast literature on DOA estimation to further optimize the performance of this algorithm.

We provide numerical results demonstrating one particular DOA algorithm ESPRIT that is computationally efficient which is, to our knowledge, on par with the best known query complexity and the best known parallel query complexity by over an order of magnitude over previous approaches.

We can quantify the practical impact of the \ac{AE} algorithm introduced in this manuscript by considering the problem of financial derivative pricing \cite{rebentrost2018quantum, Stamatopoulos2020optionpricingusing, Chakrabarti2021thresholdquantum, Stamatopoulos2022towardsquantum}. In Ref.~\cite{Stamatopoulos_2024} the authors calculate that quantum advantage in pricing an autocallable derivative contract to accuracy $2 \times 10^{-3}$ with confidence $68\%$ requires a T-depth of $4.5 \times 10^{7}$, T-count of $2.4 \times 10^{9}$ and in order for the calculation to meet the classical target of $1$ second, a $45$ MHz logical clock rate. This calculation assumes the IQAE variant \cite{grinko2021iterative} is employed at $\varepsilon=10^{-3}$ and confidence $\delta=68\%$ such that the total number of oracle calls is $C/\varepsilon = 5735$ \cite{Chakrabarti2021thresholdquantum}. From the fifth column of Table \ref{tab:target_error}, we observe that the total number of oracle calls to achieve the same accuracy and confidence using csAE is $1670$, a $ \sim3.43\times$ reduction in both T-depth and T-count. Moreover, because csAE allows for parallelization, assuming it can be fully harnessed across multiple QPUs, the deepest circuit requires $102$ oracle calls (sixth column of Table \ref{tab:target_error}), such that the logical clock rate required for quantum advantage is in theory reduced from $45$ MHz to $0.8$ MHz, a factor of $56$x reduction. However, while this figure might be of consequence in a discussion of whether a QPU with the calculated specifications for advantage can be constructed, it assumes that we parallelize the quantum algorithm, but not the classical counterpart used for comparison, Monte Carlo. Because Monte Carlo can be parallelized in a straightforward manner, a fair comparison would allow for both quantum and classical methods to execute in a parallel fashion. From Fig.~\ref{fig:constant_factor_plot}, we observe that the csAE runtime scales as $\sim 1/N$ for $N$ processors, precisely the same way a classical Monte Carlo simulation scales with the number of processors. This means that no further logical clock rate benefit arises from the parallelization of csAE if we allow the same parallelization to Monte Carlo. Nevertheless, csAE does provide the possibility of parallelization, enabling trade-offs between circuit width and depth when desirable, unlike other high-performing \ac{AE} variants like IQAE and chebAE, overcoming an important limitation of \ac{AE} methods when compared to classical Monte Carlo.

There are a number of avenues to further improve or generalize our results depending on specific use cases. 
One obvious example is to use more sophisticated DOA algorithms on the signal.
Despite this, our numerical results using an ``off the shelf'' implementation of ESPRIT already is on par with the previous best published results for worst-case complexity. We leave for future work further optimizations that may lead to even better constant factors by taking into account the actual noise distribution. Results in the DOA literature suggest that this can lead to substantial improvements (see e.g. Ref.~\cite{Sengupta1994NonGaussian}). We further remark that the noise resilience of this approach may make this method well-suited to NISQ-based algorithms that require amplitude-estimation as well \cite{Wang2019vqe, Wang2021noisy}.

We chose a particular sampling strategy that we refer to as the $2q$-array. This choice, however, is not limiting and in fact there is a great degree of flexibility. We demonstrated the use of one particular $2q$-array, but note that there is a freedom here that one can exploit to further optimize the algorithm. One can also choose different array geometries such as co-prime arrays \cite{pal2011coprime}, ruler arrays \cite{Siavash2012ruler}, fractal arrays \cite{Regev2020fractalarrays}, along with many other approaches \cite{electronics11203334, Aboumahmoud2021sparsereview}. Different array choices can lead to different asymptotic complexity $\mathcal{O}(1/\varepsilon^\alpha)$ for various values of $\alpha$ allowing one to achieve lower circuit depth at the cost of needing more samples. This can be useful for noisy processors where circuit depth is limited \cite{Wang2019vqe, Wang2021noisy, giurgica2022low}.

We hope that this identification of quantum amplitude estimation as a special case of the well-studied direction of arrival estimation problem opens the door to many further improvements and generalizations.

\acknowledgments
We thank Peter Johnson for constructive feedback on this manuscript and Patrick Rall, Bryce Fuller and Stefan Woerner for discussions on amplitude estimation. We also thank Ethan Davies and Pei Zeng for finding a (serious) mistake in the previous version of the paper.

\bibliography{ae-algo}

\clearpage
\onecolumngrid

\appendix
\section{Array Parameters}\label{app:target_error}
Here we report detailed values for the array parameters used to generate the results presented in Tab.~\ref{tab:target_error}.

\begin{table*}[h]
\caption{\label{tab:array_params99}Parameter values for $\varepsilon_{0.99} \le 10^{-3}$.}
\begin{ruledtabular}
\begin{tabular}{rc}
Simulation parameters &$K=8.1$ \\
Array Parameters & $[2, 2, 2, 2, 2, 2, 2, 2, 2]$  \\
Query Depths     & $[0,   1,   2,   4,   8,  16,  32,  64, 128, 256]$ \\
Shots per Depth  & $[162, 73, 65, 57, 49, 41, 33, 25, 17, 9]$  \\
Total Queries    & $8,777$ \\
Max Depth        & $256$
\end{tabular}
\end{ruledtabular}

\caption{\label{tab:array_params95}Parameter values for $\varepsilon_{0.95} \le 10^{-3}$.}
\begin{ruledtabular}
\begin{tabular}{rc}
Simulation parameters & $K=4$ \\
Array Parameters & $[2, 2, 4, 2, 2, 2, 2, 2]$  \\
Query Depths     & $[  0,   1,   2,   4,   8,  12,  16,  32,  64, 128, 256]$ \\
Shots per Depth  & $[88, 40, 36, 32, 28, 24, 20, 16, 12, 8, 4]$  \\
Total Queries    & $4,488$ \\
Max Depth        & $256$
\end{tabular}
\end{ruledtabular}

\caption{\label{tab:array_params68}Parameter values for $\varepsilon_{0.68} \le 10^{-3}$.}
\begin{ruledtabular}
\begin{tabular}{rc}
Simulation parameters & $K=3$ \\
Array Parameters & $[2, 2, 2, 2, 2, 2, 2, 2]$  \\
Query Depths     & $[  0,   1,   2,   4,   8,  16,  32,  64, 128]$ \\
Shots per Depth  & $[54, 24, 21, 18, 15, 12, 9, 6, 3]$ \\
Total Queries    &  $1,560$ \\
Max Depth        & $128$
\end{tabular}
\end{ruledtabular}
\end{table*}

\end{document}